\documentclass[12pt]{article}
\usepackage{graphicx}
\newcommand{\beq}{\begin{equation}}
\newcommand{\eeq}{\end{equation}}
\newcommand{\bbar}{\begin{eqnarray}}
\newcommand{\eear}{\end{eqnarray}}

\normalsize

\begin{document}


\begin{center}
{\LARGE \bf Nonlinear waves}
\end{center}

\begin{center}
{\LARGE \bf in double-stranded DNA}
\end{center}

\begin{center}
{\large \bf Natalia L. Komarova$^{1,2}$ and Avy Soffer$^{1,2}$}
\end{center}

\noindent $^1$ Department of Mathematics, 
Rutgers University, 
110 Frelinghuysen Rd.,
Piscataway, NJ 08854S

\noindent $^2$Institute for Advanced Study, Einstein Dr., 
Princeton, NJ 08540

{\bf We propose a nonlinear model derived from first principles, to
describe bubble dynamics of DNA. Our model equations include a term
derived from the dissipative effect of intermolecular vibrational
modes. Such modes are excited by the propagating bubble, and we term
it "curvature dissipation".  The equations we derive allow for stable
pinned localized kinks which form the bubble. We perform the
stability analysis and specify the energy requirements for the motion
of the localized solutions. Our findings are consistent with
properties of DNA dynamics, and can be used in models for denaturation
bubbles, RNA and DNA transcription, nucleotide excision repair and
meiotic recombination.}

$\vspace{0.2cm}$

\noindent Keywords: Denaturation bubble, transcription, soliton, kink, dissipation.

\newpage 

\section{Introduction}
Recent experimental breakthrough works \cite{lib}, \cite{wan},
\cite{pog}, \cite{ash}, \cite{han}, \cite{koc} on DNA strand
separation in transcription, denaturation and other processes, have
made it possible to develop a detailed understanding of such
fundamental steps of life (e.g. protein production from the DNA
template, see below).  There has been a parallel extensive work on
theoretical models to describe the corresponding ``bubble'' dynamics.
One line of research concentrated on the nonlinear nature of the
denaturation opening, see e.g. \cite{eng}, \cite{fed}, \cite{yak1},
\cite{mut}, \cite{pey}, \cite{dau}, \cite{bog}, \cite{bar},
\cite{bha}, \cite{cam}. In these models, the denaturation bubble is
described either as a breather or a kink of nonlinear Klein-Gordon
(NLKG) or of sine-Gordon type equations. Statistical models to include
temperature and noise were also developed \cite{kaf}, \cite{the}. A
number of detailed numerical studies of local DNA opening have been
carried out, which are detailed enough to be compared with
experimental data on long chains of DNA strands. These include, among
others, studies of Lavery and colleagues, by methods of molecular
mechanics \cite{ram}, \cite{ber}, and molecular/Brownian dynamics
\cite{bri}, \cite{gui}; molecular dynamics simulations of DNA by the
group of Beveridge \cite{bev}, \cite{mcc}, \cite{mcc1} and Langowski
\cite{bus}, \cite{lan}. In a series of works, Schlick and her group
developed comprehensive models of DNA dynamics. This approach uses a
bead model of DNA chains, where the energy of the chain depends on
twist, stretch, bend and hydrodynamic-mediated inter-base
interactions. This system is then assumed to obey the Langevin
equation, and both inertial and overdamped cases are studied in
detail, \cite{bea1}, \cite{yan}, \cite{rama}, \cite{sch}.

The simulation of dynamics of DNA on relevant time scales of
transcription and folding is extremely computationally extensive. One
of the important features is that the relaxation times are at the
picosecond level, while we need to follow the dynamics in the
millisecond and second level, \cite{sch1}. Therefore it is desirable to
find a nonlinear equation that governs the bubble dynamics, while
taking into account all the physics, including the double stand
nature of DNA and sequence dependence, curvature, solvent effects etc. 

Our model attempts to understand the bubble in the double stranded DNA
using a classical mechanical model, which is in many ways similar to
models discussed above. We start by looking at a chain of connected
masses, and take the by-layer structure into account. Dissipation of
the molecular dynamics in fluid is usually derived from the
interactions between the DNA and the polarization and kinematic
properties of the water molecules surrounding it.  The result is a
friction force proportional to the velocity and stochastic term to
describe the effect of thermal motion.  However, this assumes that the
DNA is a point particle, with no shape and internal degrees of
freedom. The effect of internal degrees of freedom may also be
relevant.  It can be shown that coupling of a system to a {\it
conservative} big system leads to a form of dissipation for the small
system.  In fact, a simple motivating example was solved exactly by
Lamb in 1900, where he showed that coupling an oscillator to an
infinite string leads to dissipation. For a review of recent works in
this direction see \cite{sof}.  We then ask what will be the effect of
the internal vibrational and other modes of each base in the DNA.  By
modeling each such base as a string we derive the leading order
effect, and show it is dissipative, yet can not be incorporated in the
usual friction terms as it is curvature dependent. Hence we refer to
this contribution as {\it curvature dissipation}.  Just like the usual
friction it depends only on the first derivative in time.  On the
other hand it also contains the derivatives of the amplitude with respect to $x$,
the position along the chain. 

It is interesting to note that curvature effects are sometimes
relevant to friction, see e.g. \cite{lig}, \cite{wig},
\cite{gol}. However, the kind of dissipation considered in these and
other works comes from the interaction of a curved object with the
surrounding fluid. The curvature dissipation introduced in this paper
comes from the {\it internal} motions of the molecule and would enter the
equation of motion even if the motion took place in vacuum.

Our main result is an equation of motion for a double-stranded DNA
which allows for stable, pinned, localized solutions. These solutions
(a kink and an anti-kink) can be used to model denaturation bubbles in
many biological systems. At this first stage of developing a new
theory, we addressed the following questions: What are the static
parameters of denaturation bubbles (their steepness and their
longitudinal size)? How much energy does a bubble require to be
moved along the DNA chain (or, alternatively, how difficult is it to
keep it in place?) What defines the direction of motion of a bubble?
How can a bubble collapse?

This paper is organized as follows. We start by presenting several
biological systems where denaturation bubbles play an important role
(Section~\ref{bub}). In Section~\ref{der} we outline the derivation of
the new equation of motion for a double-stranded DNA. In Section
\ref{solutions} we find relevant solutions of this equation and study
their stability; we prove that our model supports {\it stable, pinned}
localized solutions. We also present numerical stability results for a
more complicated, spatially inhomogeneous system. In Section~\ref{bio}
we discuss properties of our model in the context of several
biological scenarios. We present model predictions on the size and
shape of the bubble, energetic requirements for bubble motion,
directionality of bubble motion, and bubble collapse. We identify the
parameters that have to be measured to validate the model. We also
define the relative importance of curvature dissipation.  Section
\ref{concl} is reserved for conclusions.

\section{\label{bub} Denaturation bubbles in biological systems}

Here we list several examples of biological phenomena where
denaturation bubbles are essential. It is remarkable that bubbles are
found at the very basis of life: reproduction (both mitosis and
meiosis) and protein synthesis.

\begin{figure}
  \centering \includegraphics[scale=.5]{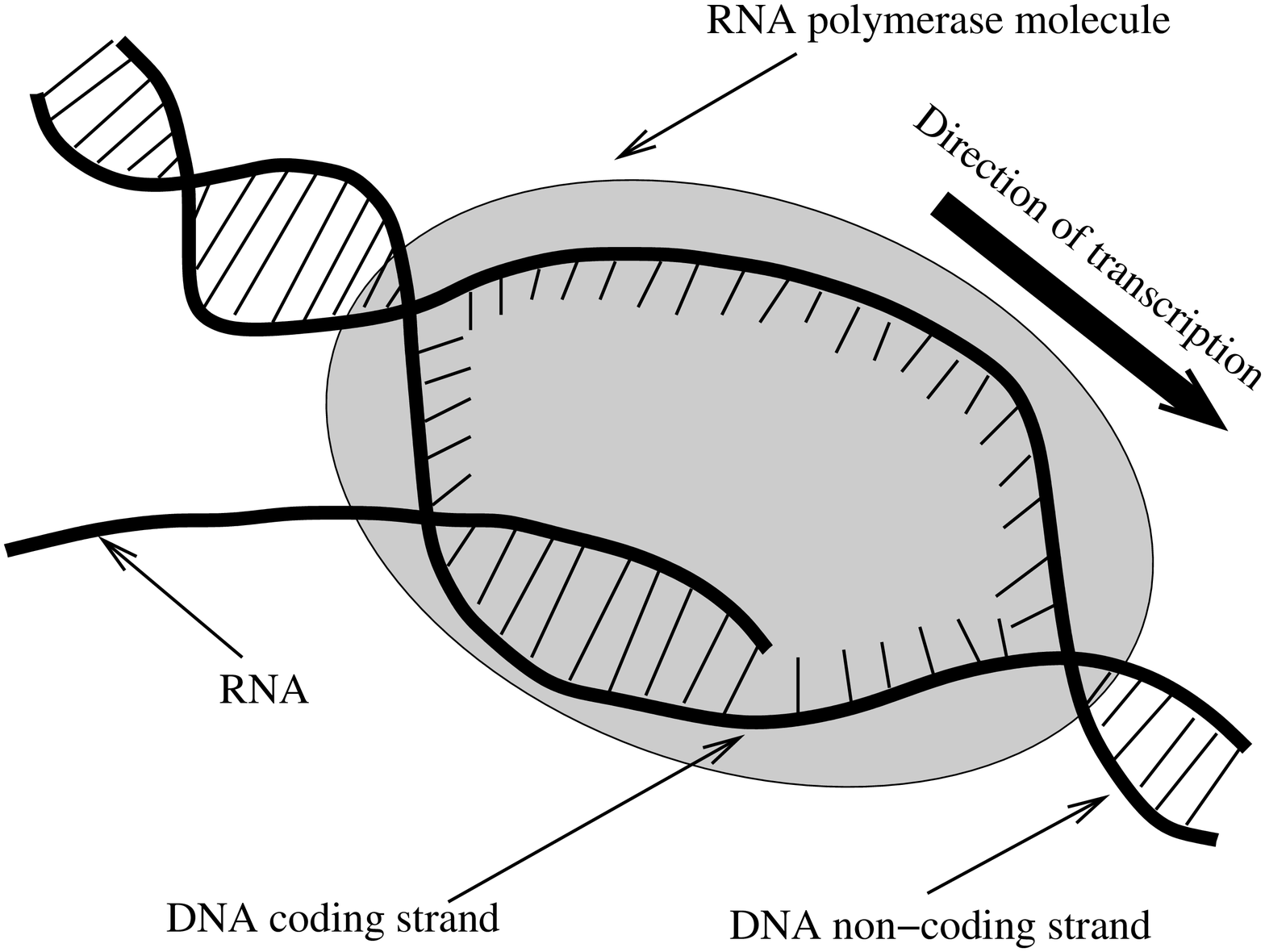}
 \caption{\footnotesize  }
\label{fig:tran}
\end{figure}

\paragraph{RNA transcription.} A denaturation bubble plays 
the central role in RNA transcription, the first step in protein
synthesis \cite{cell}, see figure \ref{fig:tran}. The process begins
when an RNA polymerase (RNAP) enzyme molecule binds to the promoter
sequence of the DNA. It starts the transcription by opening up a local
region of about 20 base-pairs on the double helix to expose the
nucleotides. One of the two strands serves as a template for
complimentary base-pairing with incoming monomers, which begin an RNA
chain. The RNAP molecule then moves stepwise along the DNA, unwinding
the DNA helix just ahead to expose a new region for base-pairing, and
rewinding the region just behind. In this process, a short region
(about 8-9 base-pairs) of DNA-RNA helix is formed briefly, after which
the newly-built region of the single-stranded DNA molecule is released
to allow the rewinding of the DNA-DNA helix. The rate of transcription
at $37$ C is about 30 nucleotides per second. A typical size of a
completed RNA chain is between $70$ and $10,000$ nucleotides.

\paragraph{Nucleotide excision repair.} A denaturation bubble plays central 
role in a repair process called {\it nucleotide excision
repair}. There, a damaged site of the DNA is recognized, and then a
bubble is formed around it, which is about $25$ \cite{eva} or $20$
\cite{mu} base pairs long for humans, and is shorter ($\sim 6$ base
pairs) in E. coli \cite{zou}. The bubble is created by a helicase
which plays a similar part at the initiation of RNA transcription. The
repair then proceeds by single strand incision at both sides of the
lesion, a removal of the damaged part from the bubble area, DNA repair
synthesis to replace the gap and ligation of the remaining single
stranded nick. 

\paragraph{Other biological systems.} A DNA bubble occurs in a variety of situations 
besides transcription and nucleotide excision repair. An expanding
bubble is formed at DNA replication. Also, we will mention the
process of meiotic recombination, where a type of a helicase (RecBCD)
propels itself along the DNA \cite{bia} creating a bubble, until a
recognition site is encountered, where the traveling loop of DNA is
cut, which initiates the genetic recombination event.

\section{\label{der}The new equation:\protect\\
motivation and derivation outline}

The motion of the double strand is usually modeled by means of some
nonlinear equation of the form, 
$$\dot z=bz_{xx}-\partial V/\partial z+\mbox{random forcing}+\mbox{higher order friction terms},$$
where $z$, the transversal displacement of the nucleotides, is a
function of space, $x$, and time, $t$, and the dot stand for its
time-derivative. The coefficient $b$ is the ``spring constant'' of the
longitudinal interactions modeled as (non)-linear springs, and $V$ is
the potential.  Different authors proposed various shapes of the
nonlinearity corresponding to the hydrogen-bond potential, $V$,
introduced nonlinearity in the ``elasticity properties'' of the
sugar-phosphate backbone, and included extra degrees of freedom coming
from the secondary structure of the DNA, as well as chiral forces and
torques. The energy terms could be very sophisticated, and often
include twist, stretch and bend of the molecule.

One common feature of nonlinear models of DNA dynamics can be
identified as follows: they rely on a somewhat {\it ad hoc} assumption that the
coupling between neighboring nodes of the lattice occurs by means of
non-material springs. The main point of this paper is to argue that
the interaction between oscillators has a different form and is better
described as coupling by ``strings'', rather than ``springs''. Roughly
speaking, we can say that some of the energy of transversal
oscillations of the double strand gets absorbed in the motion of the
material connecting the neighboring nodes. The relevant forces are
proportional to the momentum, that is, to the time-derivative of the
displacement, $z$. To leading order, this leads to a mixed-derivative
term in the master equation,
$$a(\dot{z}_{m-1}-2\dot{z}_m+\dot{z}_{m+1}).$$
The continuous version of the equation, in its simplest form, will read, 
\beq
\label{newsg}
\dot z= bz_{xx}+a\dot{z}_{xx}-\partial V/\partial z+\mbox{random 
forcing}.
\eeq
In equation (\ref{newsg}) we assume that the
constants, $a,b>0$. Here we outline the main ideas behind the
derivation of equation (\ref{newsg}). This is an equation of motion for
the node $z_m$ in the direction perpendicular to the molecule, in the
overdamped limit, where the influence of the second time-derivative
term can be neglected (see the end of Section \ref{bio} for the inertial
limit). The expression that multiplies the constant $b$ comes from the
forces acting on each node from its neighbors due to stretching (the
vibrations are not taken into account). The term $-\partial V/\partial
z$ comes from the potential forces of interaction of the two units
across the double strand. The term $\dot z$ represents the usual
friction, since the motion takes place in a viscous medium. The third
derivative term multiplying the constant $a$, the curvature
dissipation term, reflects the loss of energy due to vibrational modes
of the longitudinal connections among the nodes. In what follows, we
will outline the derivation of this new term.

In modeling double stranded molecules, one should consider the fact
that each longitudinal link is in fact a many-particle molecule, and
therefore has a large number of degrees of freedom. Such a molecule
should then be described by a dispersive system with many degrees of
freedom. In the simplest classical approximation we treat it as a
string (rather than a massless spring) of some fixed length. Note that
a more general dispersion relation than the usual string will not
change qualitatively our analysis.

\begin{figure}
  \centering \includegraphics[scale=.5]{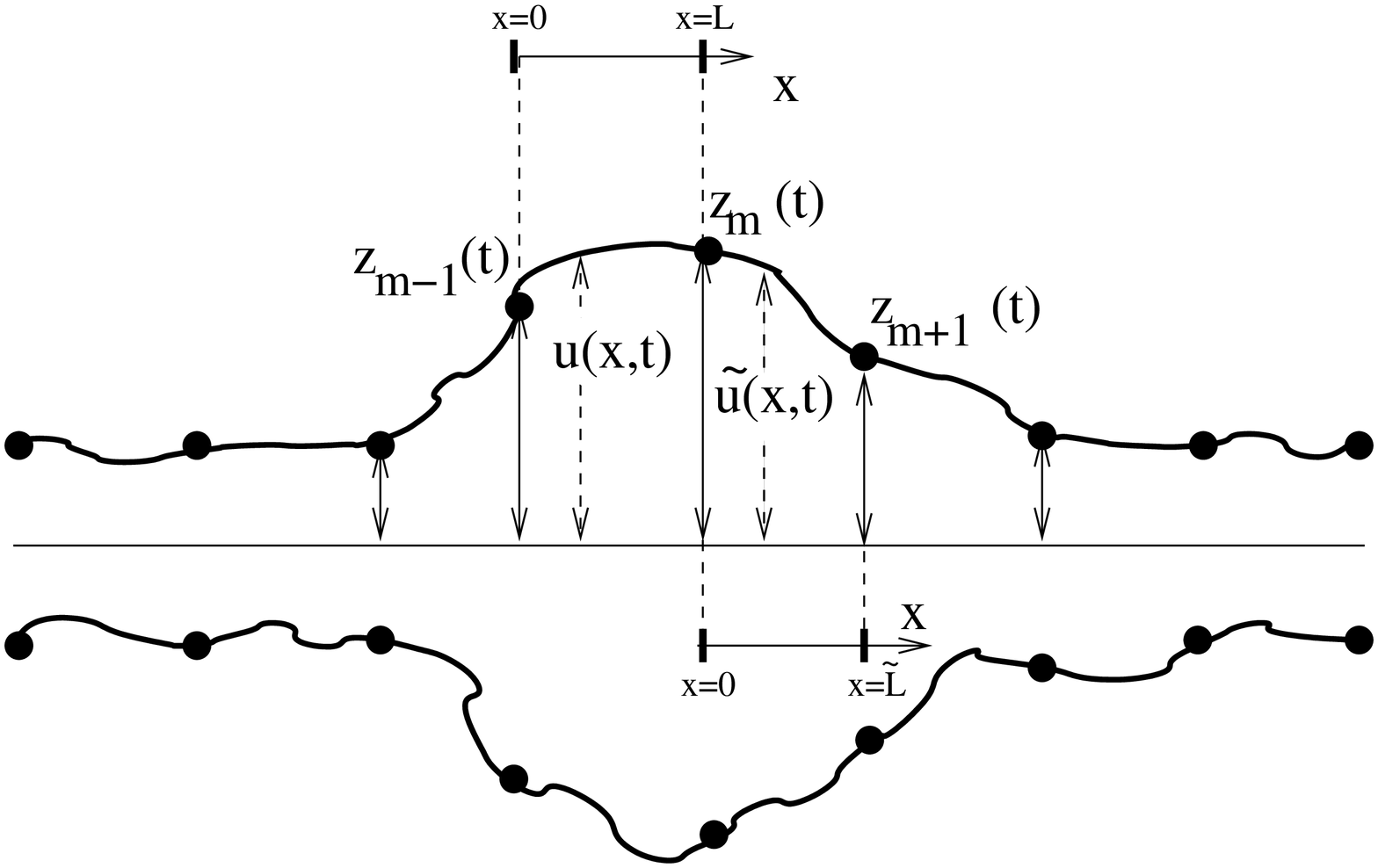}
  \caption{}\label{fig:vars}
\end{figure}

The critical difference between a ``spring'' and a ``string'' is that
a string will effectively act as a reservoir which absorbs some of the
oscillatory energy. Therefore, we expect the motion of the ends of the
links to obey an equation that contains a dissipative correction. Such
a correction can, to the leading order, be approximated by the terms
with $\dot{z}$. To see the origin of the $\dot{z}$ terms, let us
consider the node $z_m$ and solve the wave equation to the left and to
the right from it, see Fig. \ref{fig:vars}. Let $u(x,t)$ denote the
position of the string at point $x$ at time $t$. We have, to the left
of the node $z_m$,
$$u_{tt}-g^2u_{xx}=0,\quad u(0,t)=z_{m-1}(t),\quad u(L,t)=z_m(t),$$
where $L$ is the length of the connection and $g$ is the speed of
sound in the string. Similarly, to the right from the node $z_m$ we have,
$$\tilde u_{tt}-\tilde g^2\tilde u_{xx}=0,\quad \tilde u(0,t)=z_{m}(t),\quad \tilde u(\tilde L,t)=z_{m+1}(t),$$
where the constants do not have to be the same. 
%
%

The force exerted on the node from the right in the direction
perpendicular to the string is proportional to
$\tilde{u}_x\vert_{x=0}$, and the force from the left is proportional
to $u_x\vert_{x=L}$. By examining the solution of the wave
equation, we can show that the spatial and temporal derivatives are
linearly dependent; therefore, the force can be defined in terms of
$\tilde{u}_t\vert_{x=0}$ and $u_t\vert_{x=L}$. In turn, the solution
$\tilde{u}(x,t)$ is a linear functional of the boundary conditions,
$z_{m}(t)$ and $z_{m+1}(t)$, thus $\tilde{u}_t\vert_{x=0}$ is a linear
functional of $\dot{z}_{m}(t)$ and $\dot z_{m+1}(t)$, which we denote
$G(\dot{z}_{m}(t),\dot z_{m+1}(t))$. Similarly, $u_t\vert_{x=L}$ is a
linear functional of $\dot z_{m-1}(t)$ and $\dot z_{m}(t)$. Therefore,
the force from the moving springs can be expressed as
$$\tilde G(\dot{z}_{m}(t),\dot z_{m+1}(t))-G(\dot{z}_{m_1}(t),\dot z_{m}(t)).$$
In equation (\ref{newsg}) we used a very simple model for $G$ and
$\tilde G$, where they were just linear functions of their
variables. This gave rise to the term
$a(\dot{z}_{m-1}+\dot{z}_{m+1})-2a'\dot{z}_m$. Setting $a'=a$ and
taking the continuous limit, leads to equation (\ref{newsg}), which
corresponds simply to $G(y_1,y_2)=\tilde G(y_1,y_2)=a(y_2-y_1)$. A
complete derivation of the functionals $G$, $\tilde G$ will be
presented elsewhere.

\section{\label{solutions}Localized solutions and their stability}

In the literature, the DNA denaturation bubble is often modeled in
terms of kinks or breathers. However, both types of localized
solutions have several problems \cite{cam}. Breathers generically lose
stability as the level of discretization of the lattice becomes lower
\cite{aub}. Kinks, on the other hand, are very difficult to pin, even
on a lattice. As the degree of discretization decreases, the
Peierls-Nabarro barrier that keeps a kink from moving decreases
exponentially \cite{wil}, \cite{joo}. Therefore, a very small amount
of energy can set a kink in motion.

In this section, we will describe solutions of equations of type
(\ref{newsg}) which have properties relevant to many biological
systems. Namely, we will study equation 
\beq
\label{nof}
\dot z= bz_{xx}+a\dot{z}_{xx}-\partial V/\partial z,
\eeq
and prove that it supports {\it stable pinned localized solutions}.
The effects of random forcing (equation (\ref{newsg})), which is an
integral feature of dynamics on the relevant scales, has to be
analyzed separately. Stability of localized solutions in the system
without random forcing is a necessary condition for their stability
once the temperature effects have been added.

\begin{figure}
  \centering \includegraphics[scale=.5]{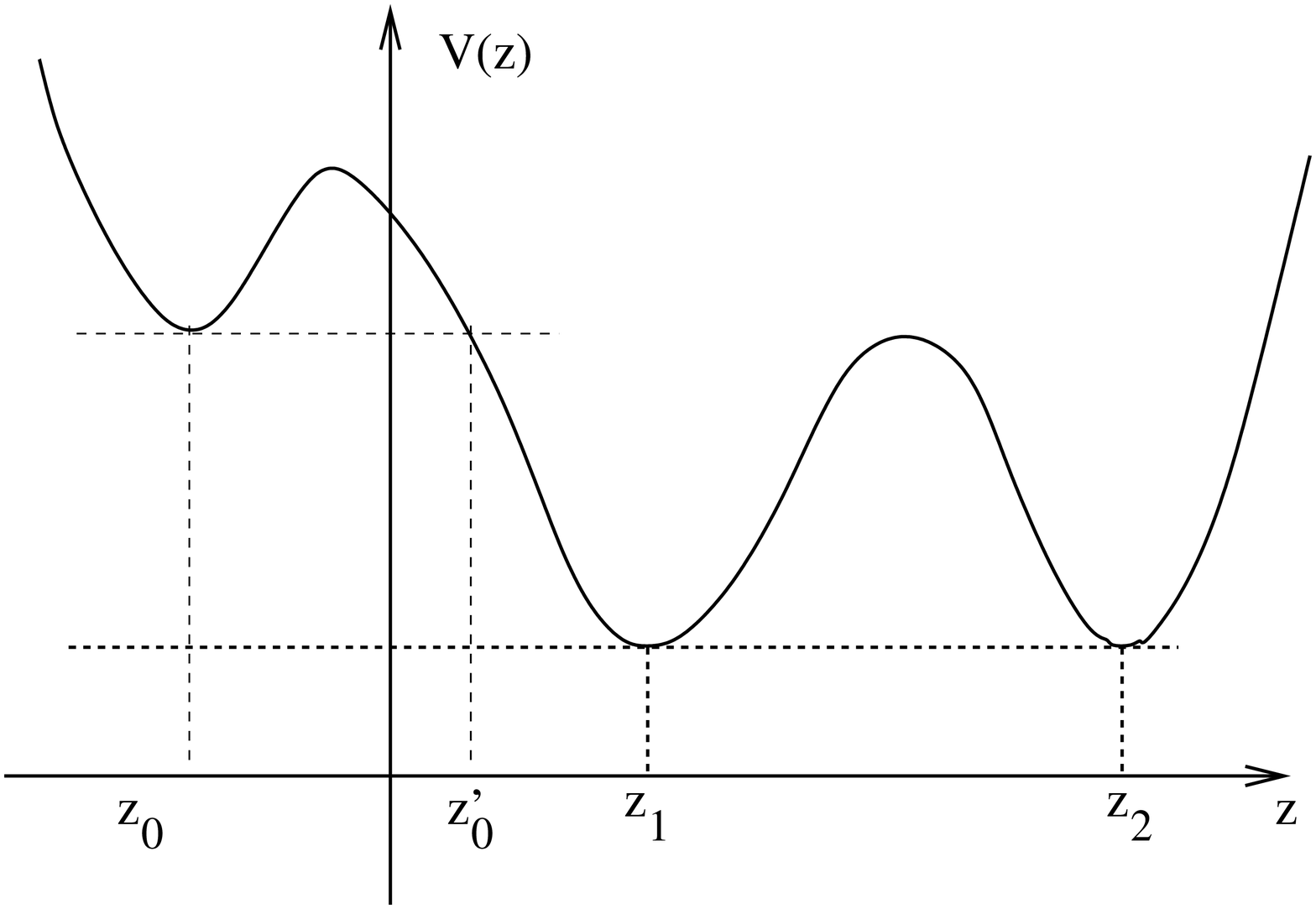}
  \caption{}\label{fig:pot}
\end{figure}

\paragraph{Kinks, solitons and the energy functional.} The exact shape of 
a stationary localized solution, $\bar z(x)$, is found from the
equation
\beq
\label{solit}
b\bar{z}_{xx}-\partial V(z)/\partial z\vert_{z=\bar z}=0.  
\eeq
Note that the nature of the solution $\bar z(x)$, will depend on the
form of the potential, $V,$ as a function of $z$. Let us suppose that
$V(z)$ is a smooth function. Integrating equation (\ref{solit}) in
$x$, we can see that the quantity $C=b\bar z_x^2/2-V(\bar z)$ is a
constant along the solution for $-\infty < x<\infty$. Using this
property, we can see that a {\it topological kink} (or antikink)
solution exists only if the potential, $V(z)$, as a function of $z$,
has at least two minima, say, at points $z_1$ and $z_2$, such that
$V(z_1)=V(z_2)$, see figure \ref{fig:pot}; here $z_1<z_2$ are some
real numbers. The kink will satisfy the conditions at infinity,
$$\lim _{x\to -\infty}\bar z(x)=z_1,\quad \lim _{x\to
+\infty}\bar z(x)=z_2.$$
For the antikink, we have
$$\lim _{x\to -\infty}\bar z(x)=z_2,\quad \lim _{x\to
+\infty}\bar z(x)=z_1.$$
A different type of localized solutions is a soliton. A soliton
solution is possible whenever the function $V(z)$ has a local minimum,
say, at a point $z=z_0$. There will be another point, $z'_0$, with
$V(z_0)=V(z'_0)$. The soliton solution will have a maximum (or
minimum) value of $z_0'$, and the following condition at infinity:
$$\lim _{x\to -\infty}\bar z(x)=\lim _{x\to
+\infty}\bar z(x)=z_0.$$

Let us define distance between functions, $z(x)$ and $v(x)$, as
$$d(z,v)=\int_{-\infty}^\infty [(z-v)^2+(z_x-v_x)^2]\,dx.$$
The energy functional of equation (\ref{nof}), is given by 
\beq
\label{energy}
E\{z\}=\int_{-\infty}^\infty \left[bz_x^2/2+V(z)-V_\infty\right]\,dx.
\eeq
Here, $V_\infty$ is some constant; we subtract this constant in order
to make sure that a localized solution has finite energy. For solitons
we take $V_\infty=V(z_0)$. For kinks, we have $V_\infty=V(z_1)$, see
figure \ref{fig:pot}. With this choice of the constant, in each case,
the function $E\{z\}$ is defined for solutions $z(x,t)$, such that
$d(z,\bar z)<\infty$, where $\bar z(x)$ is the localized
solution. Using equation (\ref{nof}), it is easy to show that
\beq
\label{decay}
\frac{dE\{z\}}{dt}=-a\int_{-\infty}^\infty z_{xt}^2\,dx-\int_{-\infty}^\infty z_{t}^2\,dx\le0,
\eeq
which means that starting from any initial conditions (for which
$E\{z\}$ is defined), the solution will always decrease the energy
functional.

\paragraph{Stability of kinks.} Here we will show that kinks are stable. The analysis 
for antikinks is similar. Let us prove that the energy functional,
$E\{z\}$, has a local minimum at the point $z=\bar z$, where $\bar z$
is a kink satisfying stationary equation (\ref{solit}). Let us
calculate the gradient and the curvature of $ E\{z\}$ at $\bar z$. We
have,
$$\left(\frac{\delta  E\{z\}}{\delta z},\psi\right)_{z=\bar z}=\int_{-\infty}^{\infty}\left(b\bar z_x\psi_x+\frac{\partial V(\bar z)}{\partial \bar z}\psi\right)\,dx=0$$
for all test functions, $\psi(x)$, in the appropriate space. Here and
below we use the short-hand notation, $\partial Q(\bar z)/\partial
\bar z\equiv \partial Q(z)/\partial z\vert_{z=\bar z}$, where $Q(z)$
is a function of $z$. Next, we evaluate
$$\left(\psi, \frac{\delta^2  E\{z\}}{\delta z^2}\psi\right)_{z=\bar z}=\int_{-\infty}^{\infty}\left(b\psi_x^2+\frac{\partial^2 v(\bar z)}{\partial \bar z^2}\psi^2\right)\,dx=\int_{-\infty}^{\infty}\left(\psi,H\psi\right)\,dx,$$
where the self-adjoint operator $H$ is given by
$$H=-b\frac{\partial^2}{\partial x^2}+\frac{\partial^2V(\bar z)}{\partial \bar z^2}.$$
Using Weyl's theorem, it is easy to show that this operator has a
positive continuous spectrum. Indeed, we have 
$$\lim_{x\to \infty}\frac{\partial^2 V(\bar z)}{\partial \bar z^2}=V''(z_1)>0,\quad \lim_{x\to -\infty}\frac{\partial^2 V(\bar z)}{\partial \bar z^2}=V''(z_2)>0, $$
that is, for large values of $x$, the potential $V''$ approaches the
value of the curvature at its minima, see figure \ref{fig:pot}. The
continuous spectrum must therefore be positive. The only negative
contribution could come from the discrete spectrum. In order to
exclude this possibility, let us consider the eigenfunction $\bar
z_x$, corresponding to the horizontal translation of the kink. This
eigenvector corresponds to the eigenvalue zero. Indeed,
differentiating equation (\ref{solit}) in $x$, we obtain $H\bar
z_x=0$. On the other hand, this eigenfunction is positive for
monotonically increasing kinks. Using Sturm's oscillation theorem we
conclude that this eigenfunction is the ground state of the operator,
which means that all other localized solutions, if they exist, have
nonnegative eigenvalues. Therefore, $(\psi,H\psi)\ge 0$.

We have proved that the function $\bar z$ is a local minimum of the
energy functional $E\{z\}$. Therefore, starting from any solution in a
vicinity of the kink, the system will return to the kink. This
concludes the stability analysis.

\paragraph{Instability of solitons.} The above argument breaks down in the case 
of solitons. It will remain the same up to the point where we look at
the second derivative of $ E\{z\}$ at the point $z=\bar z(x)$,
the soliton solution. The operator $H$ satisfies,
$$\lim_{|x|\to \infty}\frac{\partial^2 V(\bar z)}{\partial \bar z^2}=V''(z_0)>0,$$
so the continuous spectrum is positive. The minimum of $H(x)$ is
negative, because between the points $z_0$ and $z'_0$ there must be a
point, $z_*$ such that $V''(z_*)<0$, see figure \ref{fig:pot}. The
discrete spectrum must have a negative eigenvalue, because the
translational mode with a zero eigenvalue, $\bar z_x$, is not a
positive function in the case of a soliton. Therefore, by Sturm's
oscillation theorem, there will be another eigenfunction (the ground
state) with an eigenvalue between $V''(z_*)<0$ and zero. This suggests
that the soliton solution is a saddle point for the energy
functional. An infinitesimal perturbation in the ``right'' direction
will destabilize the solution and bring the system to a different
stationary state, with a lower energy, e.g. the solution
$z(x)=z_0=const$.

\paragraph{Nonhomogeneous chains: numerical stability results.} In the model discussed so far, we treated all base-pairs as if they were identical. A more accurate 
model for a double-stranded DNA will distinguish between two types of
hydrogen bonds, A--T and G--C. The two bonds are characterized by
different potentials, namely, $D_{A-T}=0.05\, eV$ and $D_{G-C}=0.075\,
eV$. In order to model a non-homogeneous DNA sequence, we can use a version of a
discretized equation,
\beq
\label{discrete}
\dot z_m=b(z_{m-1}-2z_m+z_{m+1})+a(\dot z_{m-1}-2\dot z_m+\dot z_{m+1})-\frac{\partial V_m(z)}{\partial z}\bigg\vert_{z=z_m},
\eeq
where $z_m(t)$ is the vertical displacement of each nucleotide, and
instead of one potential $V(z)$, we have functions $V_m(z)$ which
represent interaction between two nucleotides in each base pair. The
interaction potentials are allowed to differ from site to site.

We have performed numerical experiments where the functions $V_m\in
\{V_{A-T},V_{G-C}\}$ were taken from some distribution. It appears
that the stability properties of the kink are not affected by this
type of randomness as long as the values $V_{A-T}$ and $D_{G-C}$ are
not too far apart.

\paragraph{Inertial systems.} Finally, we consider systems where the motion 
is not overdamped, such as DNA molecules in non-soluble media. In this
situation we need to include the kinematic terms corresponding to
elastic modes, $\ddot z$.  Such cases arise in many applications
\cite{por}, including nanowires made of DNA strands, DNA on dry
surfaces, DNA held by electric fields and other nanodevices. In these
cases, the propagation of bubbles is governed by the following
equation:
\beq
\label{kap}
\kappa \ddot{z}=a\dot{z}_{xx}+bz_{xx}-\dot z-{\partial \tilde V}/{\partial z}+f(x,t),
\eeq
where $f(t)$ is the external force corresponding to the tweezers,
electric/magnetic fields etc.  Our stability
analysis holds almost exactly as before. In the absence of $f(x,t)$, we need to introduce the energy functional, 
$$\tilde E\{z\}=\int_{-\infty}^\infty \left[\frac{1}{2}\left(\kappa z_t^2+bz_x^2\right)+V(z)-V_\infty\right]\,dx.$$
Let $\bar z$ be the stationary localized solution, as before. It is
clear that if $\bar z$ is a local minimum of $E\{z\}$, then it is also
a local minimum of $\tilde E\{z\}$, since $z_t^2\ge 0$ and $\bar
z_t=0$. Therefore, the stability results for the kink hold in this
case.

The potential term in the absence of RNA polymerase is different, and
a bubble life-time analysis can be performed using equation
(\ref{kap}). In this case, the lifetime may be nonsmall, due to the
absence of stochastic noise.

\section{\label{bio}Biological applications}

The equation of motion derived here, with a curvature dissipation
term, can serve as a starting point to design detailed models of many
biological systems where a denaturation bubble plays a role, see
Section \ref{bub}. However, this is not the goal of the present
paper. In fact, at this stage we are still quite far from grasping all
the features of such complex biological phenomena as RNA
transcription, or nucleotide excision repair. For example, in order to
describe RNA transcription, a model must contain information on the
RNA polymerase molecule. In the present work we are mostly concerned
with properties of the double-stranded DNA molecule. A natural
question is, what is the value of this modeling for studies of real
biological systems?

We will answer this question by using the following analogy. Let us
suppose that we need to model a cruise ship. In order to accomplish
this task, we need to include all the details of the ship's
design. However, no such model would be any good unless we understand
the basic properties of water!  So a reasonable start for modeling a cruise
ship is good old fluid mechanics. Would water be able to hold a
massive object without sinking it? Can a (generic heavy) object move
along in water? How much energy does such motion take? And so on. In
the case of modeling RNA transcription, we first need to understand
how double-stranded DNA moves, and how a denaturation bubble forms,
before we can begin talking about details of the transcription process
itself. The model developed here addresses the following questions:
What is the generic shape of a denaturation bubble? What is its size?
Can a denaturation bubble be stable? How much energy does it take to
move it along the DNA molecule? How can a bubble collapse?

\begin{figure}
  \centering \includegraphics[scale=.5]{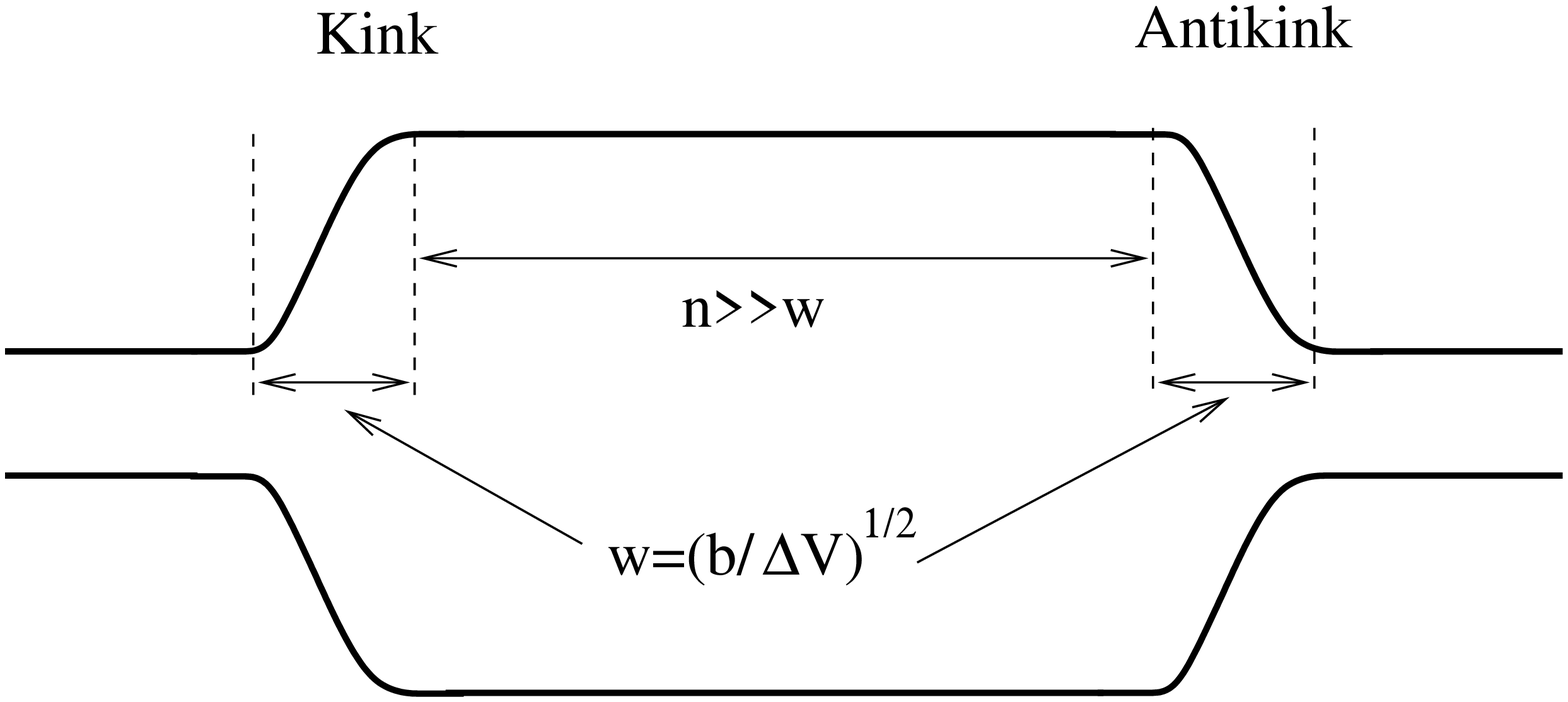}
 \caption{\footnotesize  }
\label{fig:kink}
\end{figure}

\paragraph{Shape and size of the bubble.} A denaturation bubble can be modeled as a 
solution of (\ref{newsg}) which consists of a kink and an antikink,
see Fig. \ref{fig:kink}. If the kink and the antikink are sufficiently far
apart, we can say that they do not interact and can coexist for a long
time. The width of a kink is roughly given by 
\beq
\label{ww}
w=\sqrt{b/\Delta V},
\eeq
where $\Delta V$ is the potential barrier of the interaction energy of
nucleotides across the double strand, given by the difference between
$V(z)$ at its maximum, and at its minimum. 

The longitudinal size of the bubble, $n$, is given by the distance between the
kink and the antikink. We must require that
\beq
\label{size}
n\gg \sqrt{b/\Delta V}
\eeq
in order for the bubble to be stable, see Fig. \ref{fig:kink}. Note
that the size of the bubble in this model is not defined by intrinsic
properties of the DNA molecule (except for the constraint that a
bubble cannot be too small, to satisfy condition (\ref{size})
above). This means that the bubble size can be different under
different circumstances. For instance, in RNA transcription process it
is defined by the RNA polymerase molecule. The size of the bubble
created in the process of nucleotide excision repair is defined by the
appropriate helicase. Finally, the denaturation region formed during
DNA replication or meiotic recombination does not have a fixed size,
as it is created by a moving helicase which opens up the DNA double
helix on one side of the bubble. This suggests that modeling
denaturation bubbles as a pair of two independent localized solutions
(the kink and the antikink of Fig. \ref{fig:kink}) is consistent with
biological reality, more so than using one localized solution like a
breather or a soliton.

In order to relate the model's prediction, equation (\ref{ww}), to
biological systems, we need to know numerical values for $b$ and
$\Delta V$.

\paragraph{Measurements of ``static'' parameters of the bubble.} The quantities 
relevant for the shape of the bubble (formula (\ref{ww})) are given by
$$b=\frac{Kh^2}{D},\quad \Delta V=\alpha^2h^2,$$
where $D$ is the depth of the hydrogen bond potential, $h$ is the
longitudinal distance between nucleotide pairs, $\alpha$ is the width
of the potential well and $K$ is the ``spring constant'' of the DNA
sugar-phosphate backbone. The first three parameters can be measured
relatively accurately, whereas $K$ presents a problem.

The depth of the hydrogen bond potential, $D$, has
been estimated to be $D_{A-T}=0.05\, eV$ and $D_{G-C}=0.075\, eV$ for
the two types of pairing. The parameter $\alpha$ that defines the
width of the potential well is taken to be $\alpha=2.55\,A$ in
\cite{pey}, $\alpha=4.45\,A^{-1}$ in \cite{bar}, $\alpha=4\,A^{-1}$ in
\cite{cam}. The distance between pairs is $h=3.4\,A$ (\cite{cam} and
\cite{bar}).

A more difficult quantity to measure is the ``spring constant'' $K$,
of the DNA sugar-phosphate backbone.\footnote{Large discrepancies in
the values of the spring constant are not surprising. Our models
suggests that the ``spring'' properties of the DNA, that is, the
coefficient $b$ in equation (\ref{newsg}), is not the entire
story. Energy losses due to vibrational modes of the nucleotides have
to be taken into account, which can in principle be done by measuring
the spectrum of vibrational modes.  } In \cite{pey} it was merely
estimated from the model to give a realistic denaturation temperature;
the corresponding value is $3.0\times 10^{-3}\,eV/A^2$. However, the
paper by \cite{kam} suggests that this value is much larger, the
measured parameter is $K=0.22 \,eV/A^2$. An even larger value,
$K=1.0eV/A^2$, is quoted in \cite{bar}. The paper by \cite{ger} uses
the value $K=0.026\, eV/A^2$ (however, this value has been estimated
for RNA and includes effects of the secondary structure). Note that
other experimental measurements give very different values, see
\cite{smi}, \cite{ben}, where the spring constant is found to be very
small, of the order of $10^{-6}\, eV/A^2$. However, it must be noted
that in those experiments the spring constant of the DNA molecule as a
whole was measured as opposed to local elasticity properties of the
sugar-phosphate backbone, and it is the latter quantity which is of
interest to us.

With the information that we have so far, we can obtain the value of
$w$ between $0.18$ (for $K=0.2\, eV$, $\alpha=4.45\, A$ and $D=0.33\,
eV$) and $1.96$ (for $K=1\,eV$, $\alpha=2.55$ and $D=0.4\,eV$). This
means that the number of nodes in the ``knee'' of the kink is of the
order one. This estimate is consistent with the picture of RNA
transcription (Section \ref{bub}) where an RNA polymerase enzyme
molecule opens up only a few base-pairs to complete the transcription
of a small portion of the DNA template, with 3 or fewer nucleotides
forming the ``sides'' of the bubble.

\paragraph{Energy needed to move the bubble.} In potential systems,
such as nonlinear Klein-Gordon equation, a whole family of moving
kinks, $\bar{z}(x-vt)$, exists for any velocity $v$. Therefore, moving
a kink along a lattice does not take any energy. In the new equation,
this is not the case. In order to move the bubble along the DNA
molecule, an external force must be applied. 

This has relevance for many biological systems involving denaturation
bubbles. In the context of RNA polymerase, we can ask: how strong a
push does a transcription bubble need to travel along the DNA? The
RNAP molecule is thought to be a molecular motor, which uses the
energy of ribonucleoside triphosphades to propel itself in the 3'-5'
direction along the coding strand of the DNA molecule \cite{gel}. Our
model implies that the RNAP ``drags'' the transcription bubble
(consisting of a kink and and antikink) along, using the appropriate
fraction of its total energy. The same holds for expanding
denaturation regions during the process of DNA replication and meiotic
recombination. How much energy does a helicase need to propel a
traveling loop of DNA? 

In the context of nucleotide excision repair, one can ask the
opposite question: how stable is the bubble? How easy is it to keep it
in place for as long as it takes to perform the repair?

Theoretically we can address these questions in the framework of our
model. Using equation (\ref{decay}), we can calculate how much energy
it takes to move a kink with velocity $v$ for time $\Delta t$:
\beq
\label{ener}
\Delta E=v^2\Delta t\int_{-\infty}^\infty \left(\bar z_x^2+a\bar z_{xx}^2\right)\,dx.
\eeq
We can see that energy losses come from two sources: the first term
under the integral is the usual dissipation. The second term is the
curvature dissipation, that is, the loss due to internal vibrational
modes of the DNA molecule. This is the novel contribution of the
present model.

In order to obtain a quantitative prediction, several detailed
measurements must be performed. First of all, the shape of the bubble
has to be identified, to find the slope, $z_x$ and the curvature,
$z_{xx}$ along the bubble. Then, the velocity of motion, $v$, has to
be estimated during a time-interval, $\Delta t$. Finally, the
contribution of dissipation and curvature dissipation must be
identified. This is the most difficult task.  Measuring the spectrum
of vibrational modes will eventually lead to the information necessary
to estimate the coefficient $a$ in equations (\ref{newsg}) and
(\ref{ener}). ?? AVY ADD SOMETHING ??

\paragraph{Direction of the bubble motion.} According to our
model, the bubble motion direction is defined externally, by the
``motor'' which propels the kink along the DNA chain. In biological
systems, bubble motion happens in a fixed direction. For example, in
RNA transcription, the process of elongation always proceeds in the
5'--3' direction (i.e. the RNA polymerase moves along the template
strand of DNA in the 3'--5' direction). Therefore, the ``polarity'' of
the coding strand defines the arrow of motion. Our suggestion is that
it is the molecular motor (RNAP) that recognizes the directionality of
the DNA strand, and the bubble itself can be moved in either
direction.

\paragraph{Relative importance of curvature dissipation.} Recent measurements of 
both intermolecular and intramolecular vibrational modes of
nucleotides show their significance for DNA dynamics.  In the works of
\cite{lee}, \cite{lee2}, the Raman spectrum of nucleotides is measured
in the range from $200$ to $4000$ cm$^{-1}$. These modes correspond to
the internal vibrations within the molecule, and they are in the same
energy range as the hydrogen bonds between the strands. Moreover, when
the measurements are done at low temperature ($10$--$20$K) one
observes that broad absorption lines are in fact many resonances,
fused together due to thermal fluctuations. Other modes correspond to
vibrations of two coupled nucleotides; they have a lower energy and
therefore are easier to excite. These modes have also been measured
and are typically in the range from $30$ to $150$cm$^{-1}$. Some of
them are measured by \cite{fis} in experiments on crystals, and by
\cite{bol} on the molecular level; see also \cite{wilma}.  It is clear
from this abundance of the modes at the relevant energy scales, that a
realistic temporal description, as required for example for bubble
motion in transcription, must adequately incorporate the corresponding
contributions.

We can use simple energetic considerations to estimate the effect of
curvature dissipation on the DNA dynamics.  As the bubble propagates
through the DNA, it excites many internal modes. Therefore, we need to
compare the energies of the motion of the bubble with the vibrational
modes.  The number of relevant modes of one nucleotide is multiplied
by the number of points where the curvature is not zero, about 6 (that
is, $12$ nucleotides). Using the fact that each vibrational mode of
nucleotides, as well as nucleotide-nucleotide couplings, is of the
order of $10^{-3}$ $eV$ \cite{wilma}, we can see that each vibrational
mode of the 6 involved base-pairs contributes an amount which is about
10-20\% of the difference of the base-coupling energies ($0.25\,eV$).

It is now possible to estimate the value of the coefficient $a$ in
equation (\ref{newsg}). Let us suppose that the bubble moves with a constant
speed, $v$.  Then, we assume that it excites $12$ nucleotides, with the total
energy $e$.  Then the energy change per unit time is given by $ev/d$, where
$d$ is the size of the kink (say, $d=3$ base pairs). Our formula for the rate of energy change related to internal modes is given by $av^2\int\bar z_{xx}^2\,dx$. 
Therefore
$$a\sim \frac{e}{vd\int \bar z_{xx}^2\,dx},$$
where the integral is completely determined by the shape of the kink
and has support ($\sim d$) of a few base points.  Now, notice that we
can eliminate the dependence on $v$ using the fact that E, the total
energy is proportional to $ v^2$. Therefore $\Delta E /E =ev/dE \sim a
$ where $ \Delta E$ stands for the energy loss per unit time.
?? AVY PLEASE CHECK ??

\paragraph{Bubble collapse.} Our approach can find applications in modeling transcription
termination. There are two ways in which transcription is
terminated. $\rho$-independent termination involves a specific
sequence prone to forming a hairpin. $\rho$-dependent termination
requires a subunit of RNAP which utilizes the energy of ATP to stop
the transcription. In order to model this, we can use equation
(\ref{discrete}), a discrete version of the equation of motion where
the two different types of hydrogen bonds, A-T and G-C, are taken into
account.

The process of $\rho$-independent termination can be modeled by introducing a
large perturbation in the sequence of $V_m$ (hydrogen
bonds). Simulations show that a particularly {\it small} value of
$V_m$ at one site can lead to a collapse of the kink and the antikink
on each other. The $\rho$-dependent termination can be modeled by
adding a large perturbation somewhere between the kink and
antikink. Say, if the value of $z$ outside the bubble is $z_1$, and it
is $z_2$ inside, setting several (strategically chosen) nodes inside
the kink back to the value $z_1$ may cause a collapse of the
kink-antikink pair.

The behavior of our model is in qualitative agreement with reality. At
this stage, we can only suggest that the equations of motion that we
derived for the dynamics of a double-stranded DNA molecule allow for a
bubble collapse if appropriate forcing is applied. A more detailed
model, based on particular sequences, must be devised to give
quantitative predictions.

\section{\label{concl} Conclusions}

We have introduced a nonlinear equation of motion describing the
dynamics of double-stranded DNA. Along with the usual dissipation
term, it contains a curvature dissipation term, corresponding to the
loss of energy to the many vibrational modes of the DNA molecule. This
equation allows for a  localized, pinned solution which can
be relevant for modeling DNA denaturation bubble because of the
following useful properties:
\begin{itemize}
\item[(i)] It is not a breather, that is, its existence does not depend on the fast transversal vibrations; 
\item[(ii)] It is pinned, that is, it will not travel along the DNA when perturbed in the longitudinal direction; in fact, it requires finite energy to move;
\item[(iii)] It is stable, and its stability can be proved rigorously.  
\end{itemize}
There are many biological processes involving DNA denaturation
bubbles, such as RNA transcription, nucleotide excision repair, DNA
replication and meiotic recombination. When modeling these and other
processes, the basic equation of motion for the double-stranded
molecule must allow for stable solutions corresponding to local
opening of the DNA. In this first paper we have suggested a framework
for such modeling.

The dynamical formulation we use, allows the incorporation of other
important structural factors.  The first thing we need to include is
the effect of chain content. This is easily done by making the strength
of the coupling between the strands change value according to whether
it is GT or TC.  We can also include the effect of content by changing
the "string" constant as we move from type base to base along the
chain; for this we can use the information on relevant excited modes
of each such molecule. We note also that the effect of stacking, which
was considered by many authors before (see e.g. \cite{bar2},
\cite{bar3}) can be included in a similar way by adding "spin" degrees
of freedom. There is no reason to believe that existence and stability
results would change in the modified system.

Most importantly, the effect of curvature of the DNA molecule, can
also be implemented by making the coefficients of the discrete
Laplacian position-dependent.  It is not easy to see how this can be
done in a nonhamiltonian, energy landscape type models. The
implications of these modifications may be very important, and will be
studied in a forthcoming work. Here we only mention that the {\it
curvature effects} play a central role in the DNA dynamics, and can
have important consequences for the regulation and the dynamics of the
transcription process. For example, when the DNA region is tightly
bound, curled around a chromatin, transcription initiation is
impossible. But when the curvature is lowered, by the action of
appropriate enzymes, the process can begin. Once the process has
started, the curvature will affect the velocity of propagation of the
bubble. Most importantly, we conjecture that, in some places, it will
also change the effective energy landscape, to the point of creating,
or moving of {\it arrest points}.  Such points are critical to
understanding transcription regulation.

Finally, we would like to describe details of biochemical reaction
which include more players. In upcoming papers we will concentrate on
the process of RNA transcription and show how the equation of motion
for the double-stranded DNA can be coupled with an explicit equation
for the RNA polymerase molecule. Unfortunately, complicated models
like this do not often allow for a clear and rigorous mathematical
analysis. The advantage of the present model is its transparent
behavior. It will serve an a building block for more complicated
systems.

$\vspace{4.5cm}$

\noindent{\large \bf Figure legends}

$\vspace{0.5cm}$

\noindent Figure 1. A schematic of the RNA transcription process, showing a DNA denaturation bubble.

$\vspace{0.5cm}$

\noindent Figure 2. The geometry of the DNA molecule and the variables used to derive the equation of motion. The nodes represent the nucleotides. 

$\vspace{0.5cm}$

\noindent Figure 3. The shape of the potential which allows for a
topological kink solution and a soliton solution. Note that in our
model, the potential $V(z)$ defines interactions between nucleotides
via the hydrogen bond.  We can assume that the first minimum, $z_1$,
corresponds to the equilibrium distance between the strands of the
double helix. The other minimum, $z_2$, is one of many in a complicated
potential landscape corresponding to separated strands. In the context
of RNA transcription, the complex shape of the potential can be a
consequence of the presence of the RNAP as well as the newly-formed
RNA chain. It is not unreasonable to assume that at least one of the
minima will be sufficiently close to the minimum $V(z_1)$ to guarantee
a long-lived topological kink.

$\vspace{0.5cm}$

\noindent Figure 4. A model of a denaturation bubble as a kink-antikink pair. 

\newpage

$\vspace{10cm}$

\newpage

\newpage

$\vspace{10cm}$

\newpage

\end{document}